\documentclass[twocolumn,aps,floatfix,showpacs,bibnotes,prl]{revtex4-1}
\newcommand{\be}{\begin{equation}}
\newcommand{\ee}{\end{equation}}
\newcommand{\bea}{\begin{eqnarray}}
\newcommand{\eea}{\end{eqnarray}}

\usepackage{graphicx}
\usepackage{amsmath}
\usepackage{bm}
\usepackage{xcolor}     
\usepackage{soul}       

\begin{document}

\title{Transport of a Bose gas in 1D disordered lattices at the fluid-insulator transition}
\author{Luca Tanzi, Eleonora Lucioni, Saptarishi Chaudhuri, Lorenzo Gori, Avinash Kumar, Chiara D'Errico, Massimo Inguscio, Giovanni Modugno}
\affiliation{LENS and
Dipartimento di Fisica e Astronomia, Universit\`a di Firenze,
  and CNR-INO\\  50019 Sesto Fiorentino, Italy }

\begin{abstract}
We investigate the momentum-dependent transport of 1D quasi-condensates in quasiperiodic optical
lattices. We observe a sharp crossover from a weakly dissipative regime to a strongly unstable one
at a disorder-dependent critical momentum. In the limit of non-disordered lattices the observations
indicate a contribution of quantum phase slips to the dissipation. We identify a set of critical
disorder and interaction strengths for which such critical momentum vanishes, separating a fluid
regime from an insulating one. We relate our observation to the predicted zero-temperature
superfluid-Bose glass transition.
\end{abstract}
\pacs{03.75.Lm, 05.60.-k}

\date{\today}
\maketitle

The transport in low-dimensional superfluids and superconductors is strongly affected by the
presence of disorder, isolated defects or even a periodic lattice. The superflow tends to become
unstable for increasing velocities and decay via phase-slip nucleation, a mechanism that is
particularly strong in 1D because of the large quantum and thermal fluctuations \cite{qps}. This
mechanism is relevant for different systems, such as superfluid He in porous materials
\cite{Toda,Taniguchi,Cazalilla11}, superconducting nanowires
\cite{super1D,superreview,Khlebnikov05,Bezryadin,Astafiev} or ultracold atoms
\cite{Smerzi02,Fallani05,Fertig05,Altman05,Ketterle07,Demarco08}.

In particular, disorder has been identified as the main source of dissipation in superconductors
and superfluid He. By employing strongly-disordered nanowires close to the superconductor-insulator
transition, some degree of control of the quantum phase-slip nucleation rate was demonstrated
\cite{Astafiev}, and models of dissipation due to disorder have been developed
\cite{Khlebnikov05,Leboeuf}. A good control of the disorder is now available in ultracold atom
systems. Experiments are starting to address the open questions about the superfluid-Bose glass
transition \cite{Schulz,Fisher,Deissler10,Demarco,Schneble}, and have studied the effect of a
controlled disorder on the transport of 3D Bose-Einstein condensates \cite{Lye,Hulet}. A study of
the momentum- and disorder-dependent transport in the strongly-fluctuating 1D environment is
however still missing.

In this work we experimentally address this problem with 1D ultracold atomic bosons in
quasiperiodic lattices, which allow to simulate a controllable disorder, and tunable interaction.
We start our investigation from the limit of non-disordered lattices, where suitable theoretical
models for phase slips are available. By exciting a motion with variable momentum $p$ in systems
with relatively large density, we observe a rather sharp transition from a weakly dissipative
regime at low $p$ to a strongly unstable one at large $p$, in contrast to what was observed in
low-density systems \cite{Fertig05,Ketterle07}. Measurements of the momentum- and
interaction-dependent dissipation suggest a relevant role of quantum phase-slips. We then find that
a weak disorder tends to increase the dissipation and to reduce the critical momentum $p_c$ for the
instability. We observe that for a given interaction strength there is a critical disorder strength
above which $p_c$ vanishes, which indicates the crossover into an insulating regime. From a set of
different measurements we find a crossover line in the interaction-disorder plane that is
compatible with theoretical estimates for the superfluid-Bose glass transition at $T$=0
\cite{Lugan,Savona,Altman12,Roux}.

In the experiment we employ an ensemble of 1D quasi-condensates of $^{39}$K atoms with tunable
repulsive interaction \cite{Roati07}, moving in a harmonic trap in the presence of a quasiperiodic
lattice \cite{Fallani07}. The system is realized by splitting a 3D Bose-Einstein condensate into a
few hundreds of 1D quasi-condensates with a deep 2D lattice in the horizontal plane. Each
sub-system contains on average 50 atoms and has longitudinal (transverse) trapping frequency
$\omega_z$=2$\pi\times$150~Hz ($\omega_{\perp}$=2$\pi\times$50~kHz). Along the longitudinal
direction, a quasiperiodic lattice is created by superimposing two laser standing waves with
incommensurate wavelengths ($\lambda_1$=1064~nm, $\lambda_2$=859~nm). The first lattice is stronger
and sets the tunnelling energy $J=h\times150$~Hz, while the weaker secondary lattice sets the
amplitude $\Delta$ of the diagonal disorder \cite{Modugno09}. For $\Delta>2J$ all single-particle
eigenstates of the first lattice band are exponentially localized as in a truly disordered system
\cite{Aubry,Roati08}. The Bose-Hubbard interaction energy $U$ can be varied in the range
(0.3-10)$J$ by adjusting the atomic scattering length at a Feshbach resonance \cite{Derrico07}. The
mean atom number per site $n$, which scales approximately as $U^{-1/3}$, varies in the range of
2-4. From the width of the momentum distribution of the weakly interacting quasi-condensates we
estimate an upper limit for the average temperature of $k_B T\simeq6J$ \cite{suppl}.

\begin{figure}[htbp]
\includegraphics[width=0.9\columnwidth,clip]{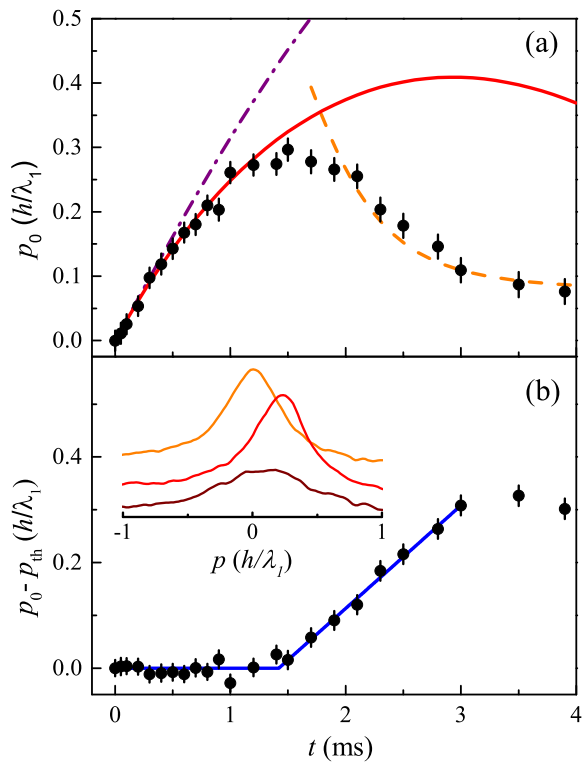}
\caption{(color online). Transport in non-disordered lattices.
a) Time-evolution of the peak momentum for $U=1.26J$ and $n=3.6$; the experimental data (dots) are fitted at short times with a
damped oscillation with $\gamma/2\pi$=135(10)Hz (continuous line) and at later times with
$\gamma'/2\pi$=600(50)Hz (dashed line). The dashed-dotted line is the expected oscillation
in absence of damping. b) The difference between the fit to the initial damped motion and
the experimental data (dots) is fitted (continuous line) to estimate the critical momentum. The inset shows
$\rho(p)$ at three different times: $t$=0, $t$=0.8ms, $t$=3.5ms, from top to bottom. The error bars represent the squared sum of
statistical and systematic uncertainties.} \label{fig1}
\end{figure}

To study the transport, the trap center along the vertical direction is suddenly displaced by a
small amount $z_0$=3.9(2)~$\mu$m by switching off a magnetic-field gradient. In absence of any
dissipation, the atoms would oscillate with a frequency $\omega^*=\omega_z \sqrt{m/m^*} \simeq
2\pi\times$90~Hz, were $m^*\simeq2.8m$ is the atomic effective mass in the lattice. After a
variable waiting time, all potentials are suddenly switched off and the momentum distribution
$\rho(p)$ is recorded after a free expansion.

We started our investigation with non-disordered lattices, i.e. $\Delta$=0, where theoretical
models are available. A typical observation of the evolution of $\rho(p)$ is shown in Fig.1, and
compared to the solution of the semiclassical equations of motion \cite{suppl}. At short times, the
displacement of the peak momentum, $p_0$, can be approximated with a damped oscillation $p_0(t)=
m^*\omega^{*2} z_0 / \omega' \sin(\omega' t) e^{-\gamma^* t}$, where
$\omega'=\sqrt{\omega^{*2}-\gamma^{*2}}$ and $\gamma^*=\gamma m/m^*$, with a damping rate $\gamma=2
\pi \times$(20-300)~Hz. At longer times, as $p_0$ increases towards the center of the Brillouin
zone ($p=h/2\lambda_1$), we observe a sudden increase of $\gamma$. This causes a stopping of the
increase of $p_0$, followed by a decay towards zero which can be again fit with a constant damping
rate of the order of 1~kHz. This change of behavior is accompanied by a sudden increase of the
width of $\rho(p)$, as shown in the inset of Fig.1, followed by a steady broadening for increasing
time.

\begin{figure}[htbp]
\includegraphics[width=0.95\columnwidth,clip]{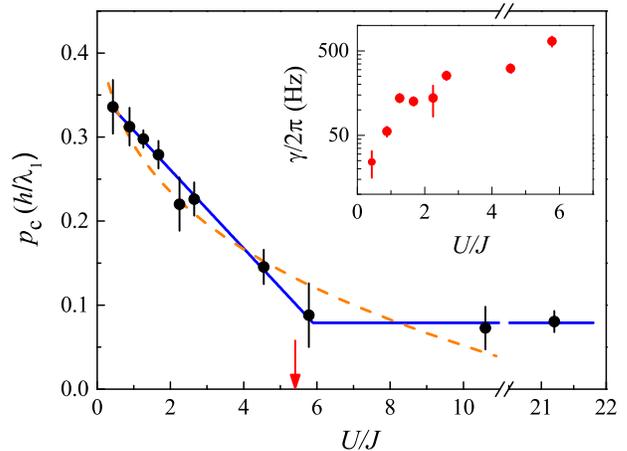}
\caption{(color online). Critical momentum for non-disordered lattices
(dots) vs the interaction energy. The continuous line is a linear fit, the
arrow marks the critical $U/J$ for the superfluid-Mott insulator transition for $n$=2,
and the dashed line is the estimated $p_c$ from the quantum phase-slips model.
Inset: initial damping rate $\gamma$.} \label{fig2}
\end{figure}

This observation is in qualitative agreement with theoretical models
\cite{Altman05,Polkovnikov05,Daley10,Polkovnikov12,Danshita13} predicting two different regimes of
quantum and thermal phase slips, in two different temperature regimes separated by a crossover
temperature $k_BT_0\simeq\sqrt{nJU}$. For $T<T_0$ quantum phase slips dominate, with an exponential
scaling of the nucleation rate with the interaction energy, density and momentum as
$\Gamma_Q\propto\exp(-7.1\sqrt{nJ/U}(\pi/2-p\lambda_1/2\hbar)^{5/2})$. For $T>T_0$, phase slips are
instead thermally activated, with a rate
$\Gamma_T\propto\exp(-4nJ/3k_BT(\pi/2-p\lambda_1/2\hbar)^3)$ \cite{Polkovnikov05}. In the framework
of these models, the weak dependence of $\gamma$ on $p$ observed in previous experiments with
low-density ($n\simeq1$) 1D bosons in lattices \cite{Fertig05,Ketterle07} was justified by the
small prefactor in the exponential scaling with $p$. Similarly, the smaller initial $\gamma$
observed in our experiment can be attributed mainly to the set of sub-systems with characteristic
$n\simeq$1. The sudden instability is instead presumably due to the higher-$n$ sub-systems,
which have a large exponential increase of $\gamma$ with $p$. We actually observe an asymmetry in
$\rho(p)$ that supports the idea of an inhomogeneous damping (inset of Fig.1). We note that our
initial $\gamma$ are comparable to those of a previous experiment in the $n<1$ regime
\cite{Fertig05}.

\begin{figure}[htbp]
\includegraphics[width=0.9\columnwidth,clip]{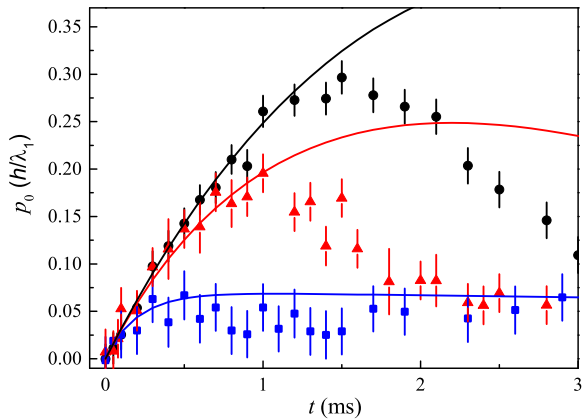}
\caption{(color online). Transport in disordered lattices.
Time-evolution of the peak momentum for $U$=1.26$J$ for $\Delta/J$=0 (dots),
$\Delta/J$=3.6 (triangles) and $\Delta/J$=10 (squares). The lines are fits
of the semiclassical motion to the initial
oscillation. The fitted damping rates are $\gamma/2\pi$=130(10)Hz,
$\gamma/2\pi$=250(30)Hz and $\gamma/2\pi$=1.1(6)kHz, respectively.} \label{fig3}
\end{figure}

We estimate a critical momentum $p_c$ separating the initial regime of weaker dissipation from the
strongly unstable regime, by linearly fitting the difference between the experiment and the fit of
the initial oscillation, as shown in Fig.1b. The measured $p_c$ features a clear decrease when
increasing $U$ at constant $J$, while $\gamma$ increases, as shown in Fig.2. Eventually, $p_c$
approaches zero as $U$ approaches the predicted critical value for the Mott insulator
($U_c/J$=2$\times$2.674 for the calculated mean occupation $n$=2 \cite{Danshita11}). Actually, even
deep into the insulating regime we observe a small but finite $p_c$ of the order of the inverse size of the
system, as already observed \cite{Ketterle07}. By a piecewise fit of the data, we obtain a critical
interaction that is comparable with theory: $U_c/J$=5.9(2)(4), where the uncertainties are
statistical and systematic, respectively. These observations lead to the conclusion
that also in 1D the onset of the Mott regime can be detected from a vanishing of $p_c$, as in 3D
systems \cite{Ketterle07}. In 1D the transport is however clearly dissipative also for $p<p_c$, as
expected.

The decrease of $p_c$ and the corresponding increase of $\gamma$ with $U$ suggest a quantum
activation of phase slip, since only $\Gamma_Q$ has a direct dependence on $U$ in the exponential.
Since phase-slips models for $\gamma$ in our large $p$ and inhomogeneous $n$ are not available, we
tentatively compare our data to the complete expressions for $\Gamma_Q$ \cite{Polkovnikov12} and
$\Gamma_T$ \cite{Polkovnikov05} regimes. In the spirit of Ref.\cite{Polkovnikov05}, we estimate
$p_c$ by imposing that the nucleation rate gets larger than the experimental damping rate
($\approx2\pi\times$1~kHz). An unknown prefactor in the calculations is adjusted to match a single
experimental data at $U/J$=4.5. The quantum phase-slip model predictions are in relatively
good agreement with the experiment, as shown in Fig.2. A similar analysis with the thermal model
predicts instead an essentially constant $p_c$ at constant $T$ (see \cite{suppl} for more
discussion). This observation is not surprising considering that $T\simeq T_0$; a careful
verification of the role of quantum and thermal phase slips is however left to future experiments
with variable $T$.

\begin{figure}[htbp]
\includegraphics[width=\columnwidth,clip]{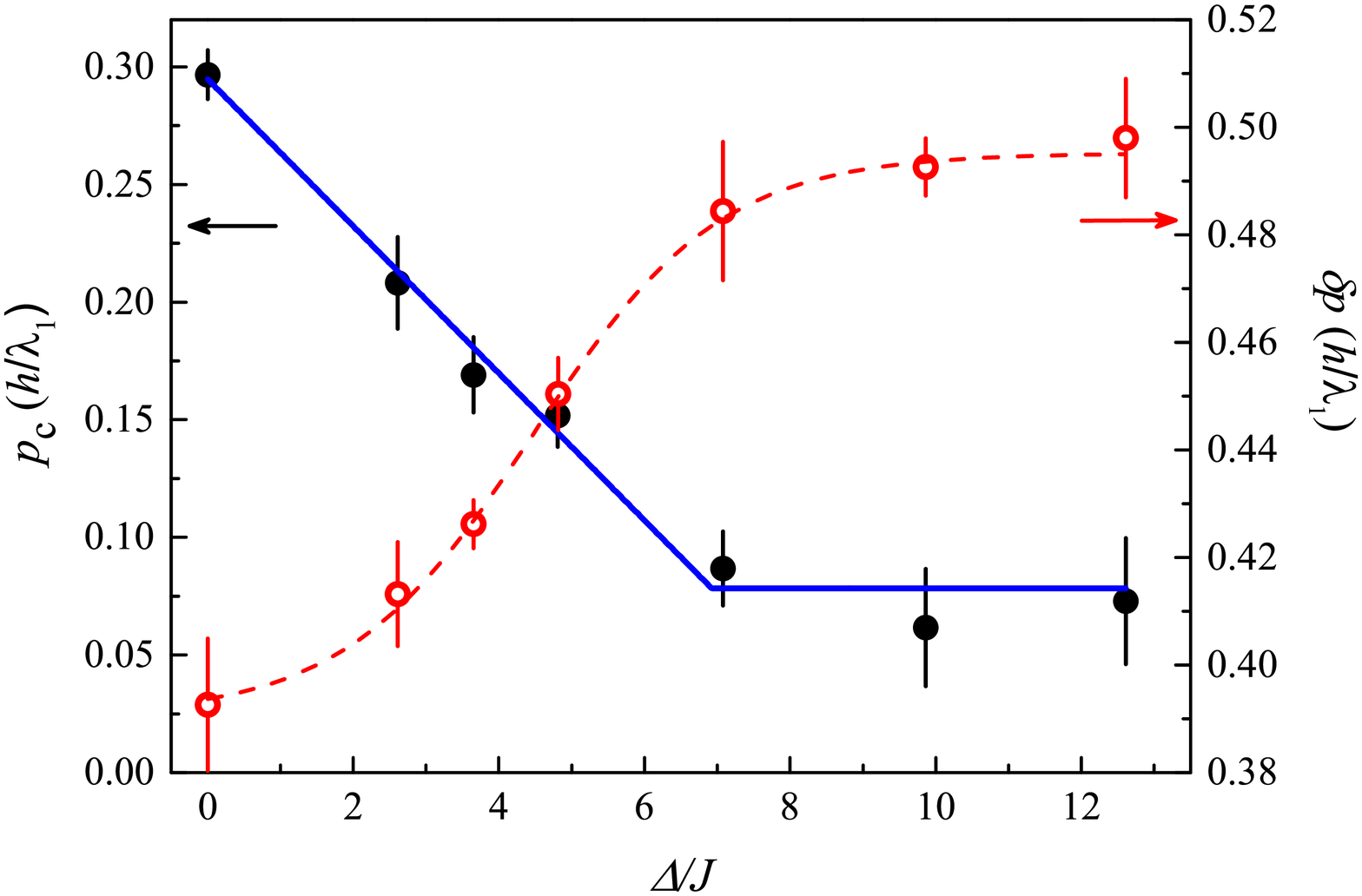}
\caption{(color online). Critical momentum $p_c$ (full circles) and initial rms momentum width $\delta p$ (open circles)for a
fixed interaction energy ($U/J$=1.26) and increasing disorder strength. A linear fit (continuous line)
 is used to estimate $\Delta_c$, while the dashed line is a sigmoidal fit of $\delta p$.} \label{fig4}
\end{figure}

Let us now turn to the transport in presence of disorder. We have in particular studied the
weakly-interacting regime, $U/J<$3, where $p_c$ for the non-disordered lattice can be very
precisely measured. The experiment is performed as before, except for a finite $\Delta$ that is
introduced together with the main lattice. Fig.3 shows how a small $\Delta$ results in a moderate
increase of $\gamma$, but also in an anticipated instability. Both changes can be related to the
idea that transport in disorder is dominated by the weakest hopping links, resulting in a smaller
effective $J(\Delta)$ that in turn produces an increase of the phase-slip nucleation rates above,
due to their exponential dependence on $J$ \cite{suppl}. A related phase-slip model developed for
disordered superconductors indicates indeed a nucleation rate scaling exponentially with $\Delta$
\cite{Khlebnikov05}, but it was derived in a different range of parameters and cannot be applied
directly to our system. An important observation shown in Fig.4 is that, for a fixed $U$, $p_c$
features a clear decreasing trend for increasing $\Delta$. Above a critical disorder strength
$\Delta_c$ of the order of the total interaction energy per atom $nU$, $p_c$ stops decreasing and
stays constant at a small value close to that observed in the Mott-insulator regime. This is
actually the regime where a weakly-interacting Bose glass is predicted to appear, since the
disorder can overcome the delocalization effect of the interaction \cite{Lugan,Savona}. The data in
Fig.4 show also that the decrease of $p_c$ is accompanied by an increase of the rms momentum width
$\delta p$ at equilibrium (i.e. at $t=0$), which is essentially the inverse of the correlation
length $\xi$, towards a saturation value. $\delta p$ starts to increase well before $p_c$ has
reached its minimum, indicating that the vanishing of $p_c$ signals the onset of a strongly
insulating phase, with a correlation length $\xi\simeq d$. Note that the observed $p$-dependent
dynamics suggests that a simpler method with a fixed observation time, used in strongly-interacting
disordered systems \cite{Demarco}, might underestimate the critical disorder strength for the
insulating regime.

Motivated by the possibility of discriminating the fluid regime from the insulating one, we have
studied how  $\Delta_c$ evolves with $U$. For each $U$, we estimated $\Delta_c$ with a piecewise
fit of the decreasing $p_c(\Delta)$, as shown in Fig.4. The summary of these measurements in Fig.5
shows a clear increase of $\Delta_c$ with $U$, indicating that the critical momentum of more
strongly interacting systems is less affected by the disorder. The increase of $\Delta_c$ is
actually fully justified, since the critical disorder strength to enter the Bose glass phase from
the superfluid in the regime of weak interactions is expected to scale as
$\Delta_c/J=A(E_{int}/J)^\alpha$, where $E_{int}\simeq nU$ is the total interaction energy per
atom, while $A$ and $\alpha$ are coefficients of the order of unity \cite{Lugan,Savona,Altman12}.
In the absence of an analytical model for the superfluid-Bose glass transition in a quasiperiodic
lattice, we fit the experimental data with $(\Delta_c-2)/J= A(nU/J)^\alpha$ to account for the
critical $\Delta/J\simeq$2 for localization in the non-interacting system. This choice is
supported by the results of the DMRG study in \cite{Roux}. The fit gives an exponent
$\alpha$=0.86(22) and a coefficient $A$=1.3(4). In the fit we excluded the data point for
$\Delta/J<2$, which should be described by a different mechanism of competition between the
miniband structure of the quasiperiodic lattice and the interaction energy \cite{Lye}.

\begin{figure}[htbp]
\includegraphics[width=0.9\columnwidth,clip]{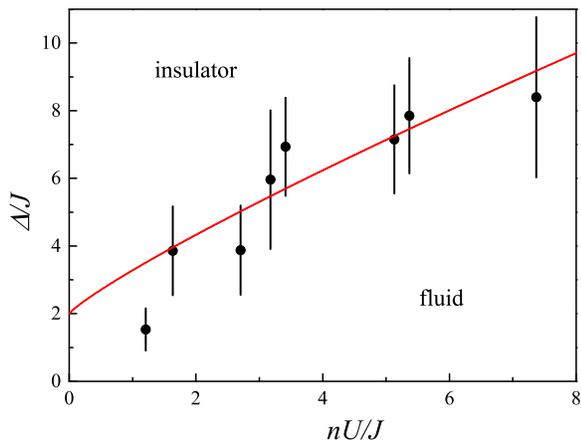}
\caption{(color online). Critical disorder to enter the insulating phase vs interaction energy.
The experimental data from the critical momentum (dots) are fitted with the model described in the text
(line).} \label{fig5}
\end{figure}

The exponent is compatible with the mean-field theory prediction $\alpha$=1 for correlated
Gaussian disorder in the so called Thomas-Fermi regime, where $E_{int}$ is larger than the typical
disorder correlation energy $E_c$ \cite{Savona}. For the quasiperiodic lattice we estimate indeed
an upper bound $E_c/J\simeq$0.7 \cite{suppl}. The observation is however not incompatible with the
prediction $\alpha=\alpha(U)<1$ found in disorder models that include corrections beyond mean-field
\cite{Altman12}. We obtain a comparable exponent, although with a different prefactor $A$, from a
similar analysis of the crossover in $\delta p$, in agreement also with previous
experiments for very small $U$ \cite{Deissler10}. It is interesting to note that many current
models for the superfluid-Bose glass transition at $T$=0 are essentially based on the evolution of
the same phase-slip nucleation rate that seems to be responsible for the dynamics observed in the
present work \cite{Altman10,Giamarchi,Prokofiev}. A careful assessment of finite-size and
finite-$T$ effects is however required to establish the relation between the observed critical line
and the theoretical fluid-insulator transition.

In conclusion, we have studied the momentum-dependent transport of 1D disordered bosons. We
have employed the vanishing of the critical momentum for the observed instability to locate the
fluid-insulator transition driven by disorder, across the interaction-disorder plane. The present
study was for weak interactions and constant $T$. Future work should explore the role of
temperature, also in connection with models for the many-body localization \cite{Aleiner}, and try
to establish a link with the Luttinger-liquid theory for the superfluid-Bose glass transition for
generic $U$ and $\Delta$ \cite{Schulz,Fisher,Altman10,Giamarchi,Prokofiev}. In this context, the
extension of the techniques used here to smaller momenta might allow to probe the predicted
universal scalings in lattices \cite{Polkovnikov12,Danshita13} and in disorder \cite{Khlebnikov05}.

We acknowledge discussions with B. Altshuler, I. Danshita, L. Pezz\'e, G. Shlyapnikov and A.
Smerzi. This work was supported by ERC (grants 203479 and 247371) and by MIUR (grants PRIN2009FBKLN
and RBFR12NLNA).

\newpage

\section{Supplementary material}

\subsection{Experimental methods and parameters}
The experiment starts with a Bose-Einstein condensate of about $N$=20000 atoms in an approximately
spherical optical dipole trap at a scattering length $a=190~a_0$. A deep 2D lattice is then
adiabatically raised in 400~ms, using S-shaped ramps. The lattice has the same spacing as the main
longitudinal lattice, $d=\lambda_1/2=532$~nm, and a typical height of 28$E_R$
($E_R=h^2/2m\lambda_1^2$ is the recoil energy), which prevents tunneling on the timescale of the
experiment. After 100~ms from the beginning of the procedure, the longitudinal quasiperiodic
lattice is adiabatically raised in 300~ms. During the last 100~ms, the scattering length is also
brought to the desired value with a linear ramp, and the optical trap is exponentially reduced to
zero.

\begin{figure}[htbp]
\includegraphics[width=0.9\columnwidth,clip]{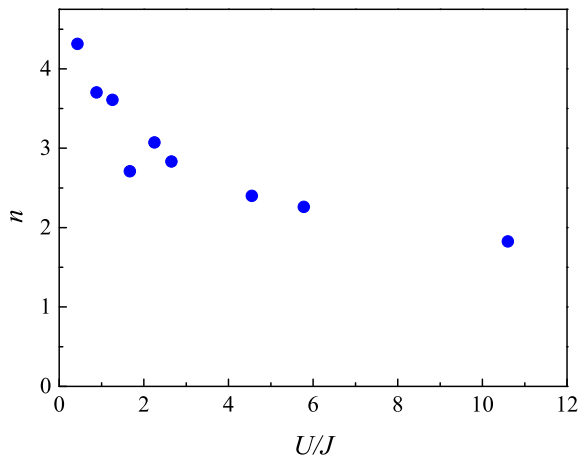}
\caption{Site occupation $n$ vs $U$ for a non-disordered lattice. $n(U)$ scales approximately as $U^{-1/3}.$} \label{fig:n}
\end{figure}

The atom number in each subsystem is estimated from the Thomas Fermi radius $R$ of the condensate
in the optical trap as
\begin{equation}
N_{i,j}= \frac{5 N d^2}{2 \pi R} \Bigl[1-\frac{(i^2+j^2)\,d^2}{R}\Bigr]\,,
\end{equation}
where the indexes ($i,j$) indicate the position with respect to the center. The mean atom number
per subsystem is about 50. The mean density for each 1D system is estimated as the largest of the
mean-field and Tonks values \cite{density}. The mean site occupation $n$ is then calculated by
averaging over all subsystems. The evolution of $n$ with $U$ in the non-disordered case is shown in
Fig.\ref{fig:n}.

The interaction energy is calculated as $U=\hbar^2/(m a_{1D})\int \varphi(z)^4 dz $, with
$\varphi(z)$ being the Gaussian approximation of the Wannier function in a lattice site. The 1D
scattering length $a_{1D}= a_{\perp}^2 (1-1.03 a/a_{\perp})/2a$, where
$a_{\perp}=\sqrt{\hbar/m\omega_{\perp}}$, can be varied by adjusting the 3D scattering length $a$
at a broad Feshbach resonance in the ground state $F$=1, $M_F$=1. The lattices are calibrated by
diffraction in either the Bragg or Raman-Nath regimes. The calibration errors on the lattices
translate in an error on $\Delta/J$ and $U/J$ of 20\% and 6\%, respectively.

The specific trap displacement $z_0$ was chosen as the smallest one allowing to reach
$p\simeq0.35h/\lambda_1$. We checked that the effect of the finite $z_0$ for disordered systems was
not significant, as discussed later.

\begin{figure}[htbp]
\includegraphics[width=0.9\columnwidth,clip]{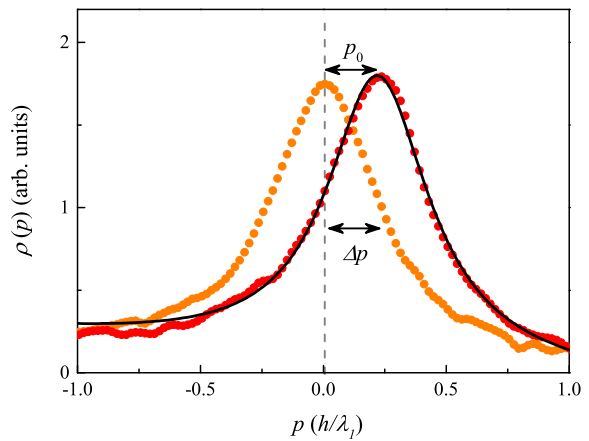}
\caption{Momentum distribution at $t$=0 (orange dots) and $t$=0.8$\mu$s (red dots), for a non-disordered lattice
and $U/J$=1.26. A fit with a Lorentzian function plus a slope (line) is used to measure the peak momentum $p_0$.} \label{fig:p}
\end{figure}

We take momentum distribution measurements by imaging the atoms after a free flight of
$t_{exp}$=16.5ms, in absence of interactions. We integrate the absorption images along the radial
direction and we analyze the resulting profiles of $\rho(p)$ which, as we discuss below, contain
also a small contribution of the in-trap position. We concentrate our analysis on the first
Brillouin zone ($|p|<h/\lambda_1$). We typically observe an evolving asymmetry of $\rho(p)$,
presumably due to the inhomogeneous damping rate. The asymmetry, which is apparent from the data in
Fig.\ref{fig:p}, leads to a systematic shift between the mean momentum and the peak momentum. In
this work we decided to study the evolution of the peak momentum $p_0$. To reduce the effect of the
experimental noise, we measure $p_0$ by fitting the data with a Lorentzian function plus a slope.

We estimate an upper bound to the mean system temperature  in the weakly-interacting,
non-disordered regime by relating the HWHM of $\rho(p)$, $\Delta p$, to a correlation length
$\xi=0.67 \hbar/\Delta p$ \cite{Gerbier}. The thermal correlation length is estimated as
$\xi_T$=1.11$\xi$, taking into account the finite width of the zero-temperature distribution. The
temperature is finally calculated as $T=\hbar^2n /\xi_T dm^*k_B$. We expect that this procedure
gives only an upper bound to the actual temperature, since there are various possible reasons for a
broadening of the observed distributions, such as a residual effect of the interaction during the
free expansion, or averaging over the different subsystems. The temperature cannot be estimated in
the strongly-correlated or disordered regimes. We however find that the entropy of the system,
which we measure by inverting the lattices loading procedure and recovering a 3D condensate, stays
approximately constant in the explored region in the $U-\Delta$ plane.

\subsection{Semiclassical analysis of the dynamics}
\begin{figure}[htbp]
\includegraphics[width=0.9\columnwidth,clip]{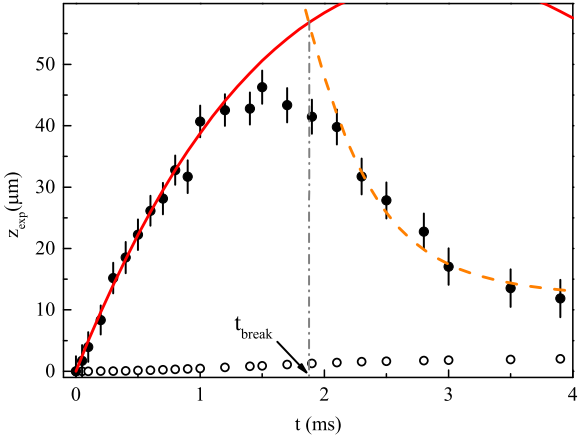}
\caption{Measured peak position $z_{exp}(t)$ after the free expansion (filled circles) and fitted in-situ position
 (empty circles) $z(t)$. The latter can be safely neglected at short times.} \label{fig:pos}
\end{figure}

The motion of the atoms in the lattice can be modeled by the solution of the semiclassical
equations of motion:
\begin{equation}\label{eq:eqofmotion}
\begin{cases}
\dot{p}= -m \omega_z^2 z -2 m \gamma \dot{z}\\
\dot{p}=m^*(p)\ddot{z}
\end{cases}
\end{equation}
with $m^{*}(p)=\hbar^2\cos(pd/\hbar)/2Jd^2$. We fit the evolution of peak of $\rho(p)$ with
$z_{exp}(t) =z(t)+p(t)t_{exp}/m$, where $z(t)$ and $k(t)$ are the solution of
Eqs.\ref{eq:eqofmotion} leaving $\gamma$ as a free fitting parameter. We note that the contribution
of the in-situ position $z(t)$ is typically rather small. For example, Fig.\ref{fig:pos} shows the
fitted $z(t)$ in comparison with the data of Fig.1 in the main paper, which is taken in absence
of disorder and for small $U$: there is a non negligible contribution of $z$ only for large times.
We can therefore quite safely neglect the contribution of $z(t)$ for short evolution times and
identify the measured distributions with $\rho(p)$.

As discussed in the main paper, we identify two different regimes in the dynamical evolution of the
system. The first part of the oscillation ($t<t_{break}$) is fitted by imposing as initial
conditions: $z(0)=z_0$, where $z_0$ is the shift of the harmonic confinement, and $\dot{z}(0)=0$.
When fitting the second part of the dynamics ($t>t_{break}$) we impose the continuity of $z_{exp}$
and of its primitive at $t=t_{break}$.

\subsection{Critical momentum, damping rates and phase-slip models for non-disordered lattices}
Although there are no theoretical models for the damping rate of the oscillations in our specific
regime of relatively large, time-dependent $p$, we can only try to compare the observations in
non-disordered lattices to existing models for the phase-slip nucleation rate. We start by noting
that the theoretical crossover temperature $k_BT_0\simeq\sqrt{nJU}$ is in the range (1.5-3.5)$J$
for our experimental parameters, hence of the same order as the estimated experimental temperature.
This suggests a possible coexistence of quantum and thermal phase slips. While the rapidly varying
$\gamma(p)$ might be consistent with both mechanisms, we find that the evolution of $p_c$ and
$\gamma$ with $U$ indicates a relevant role of quantum phase-slips.

We compare our data with the complete expressions for the phase-slip nucleation rates in the
quantum regime \cite{Polkovnikovb}
\begin{equation}
\begin{split}
\Gamma_Q=&B_Q\,L(U)\sqrt{nJU}\sqrt{\pi/2-p\lambda_1/2\hbar}\\&\sqrt{\frac{7.1(\pi/2-p\lambda_1/2\hbar)^{5/2}}
{2\pi\sqrt{U/nJ}}}\\&\exp\left[-7.1\sqrt{nJ/U}(\pi/2-p\lambda_1/2\hbar)^{5/2}\right]\,,
\end{split}
\end{equation}
and in the thermal regime \cite{Polkovnikov}
\begin{equation}
\begin{split}
 \Gamma_T=&B_T\cos(p\lambda_1/2\hbar)\sqrt{nJ/U}\\&\exp\left[-\frac{4nJ}{3k_BT}(\pi/2-p\lambda_1/2\hbar)^3\right]\,.
\end{split}
\end{equation}
Here, $L(U)\propto U^{1/3}$ is the length of the average subsystem in the Thomas-Fermi regime,
while $B_Q$ and $B_T$ are phenomenological constants that adapt the phase-slip nucleation rate to
the damping rate. A recent work actually shows that in the quantum regime $\gamma$ should be
related to $\Gamma_Q/p$, but this was so far proved only for very small $p$ \cite{Danshita}. Both
expressions are derived from a phase model that is expected to work in the regime of large $n/U$.

\begin{figure}[htbp]
\includegraphics[width=0.9\columnwidth,clip]{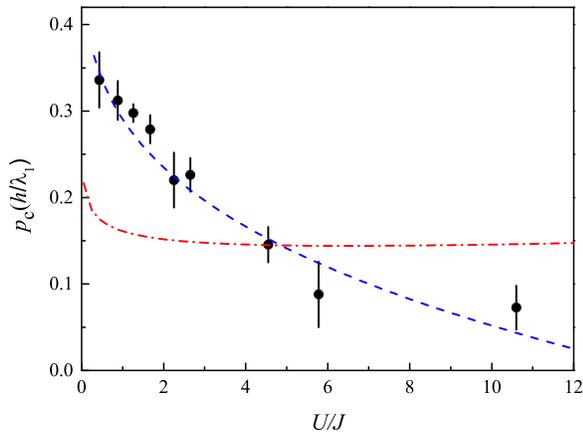}
\caption{The critical momentum for non-disordered lattices (dots) is compared with the predictions for the
phase-slip nucleation rate in the quantum (blue dashed line) and thermal (red dash-dotted line) regimes. }
\label{fig:pc}
\end{figure}

We employed these two models to estimate the critical momentum for the onset of a strongly
dissipative regime, with $\gamma/2\pi\simeq$1kHz as in the experiment. We used all the parameters
as in the experiment, $T$=6$J$, and we adjusted the arbitrary constants $B_Q$ and $B_T$ to
reproduce the observed $p_c$ at $U/J$=4.5. The experimental data and the two theoretical
predictions for $p_c$ are shown in Fig.\ref{fig:pc}. While the quantum phase-slip rate can capture
the observed evolution of $p_c (U)$, the thermal rate has only a weak dependence on $U$ at constant
temperature. This is consequence of the larger increase of $\gamma(U)$ in the quantum regime, where
$U$ enters explicitly in the exponential, than in the thermal one, where it enters only indirectly
through $n(U)$.

\begin{figure}[htbp]
\includegraphics[width=0.9\columnwidth,clip]{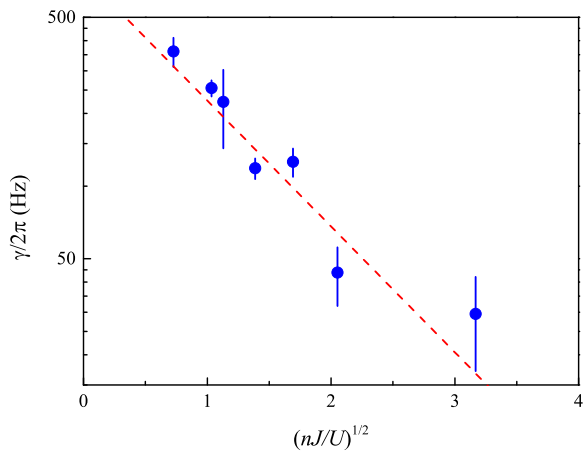}
\caption{The initial $\gamma$ vs rescaled interaction and site occupation in non-disordered lattices. The
experimental data are compared to the simplified exponential scaling described in the text (dashed line).}
\label{fig:gammaint}
\end{figure}

As for the evolution of the initial $\gamma$ with $U$, the above model cannot be easily employed
since the measurements are performed with a rapidly varying $p$. The presence of $p$ in the
exponential terms results indeed in completely different predictions for different choices of $p$.
In the spirit of the analysis performed in a previous experiment in 3D \cite{Demarcops}, we also
heuristically plotted $\log(\gamma)$ vs $\sqrt{nJ/U}$, as shown in Fig.\ref{fig:gammaint}. One
notices a surprising agreement with a $p$-independent scaling of the form
$\gamma/2\pi=a\exp(-b\sqrt{nJ/U})$, with $a\simeq$1.5~kHz and a prefactor close to unity
($b\simeq$1.2) in the exponential. We have currently no explanation for this observation.

In summary, the present measurements suggest a contribution of quantum phase slips to the observed
dissipative dynamics. So far we could not study in a reliable way the evolution of the damping
rates and the critical momenta with the temperature. In the experiment $T$ can be somehow
controlled only towards higher values, but the measurement techniques are reliable only for low
$T$. Further experiments with a controllable $T$, for example through a variation of the axial
trapping frequency, will be needed to have a true quantitative assessment of the role of quantum
and thermal phase slips.

Our measured $\gamma$ can be compared to those observed in a previous experiment with
non-disordered lattices with small density, $n\leq2$, which was performed with a similar technique
\cite{Fertig}. There, the interaction energy of a system with fixed scattering length was tuned by
changing the lattice depth. The same range of interaction energy of our experiment
$E_{int}=nU=(0.3-3)J$, with $J/h$=150Hz, was reached only for very low lattice depths, in the range
$(0.5-3)E_R$. The reported damping rates were in the range $\gamma/2\pi$=(6-600)~Hz, which is
approximately the same range of our measured rates.

\subsection{Critical momentum, damping rates and finite-displacement effects for disordered lattices}

\begin{figure}[htbp]
\includegraphics[width=0.9\columnwidth,clip]{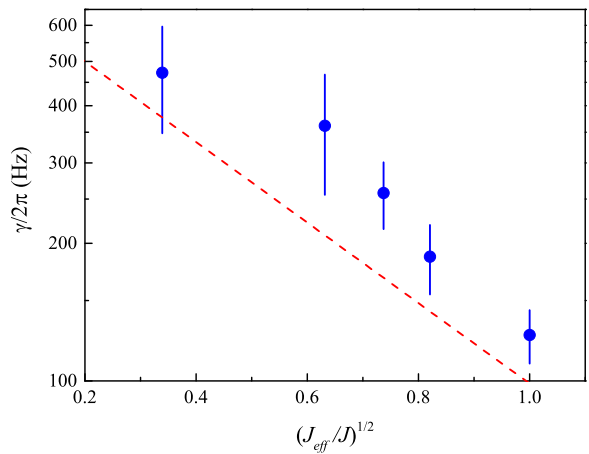}
\caption{Initial damping rate for disordered lattices vs the effective tunnelling $J_{eff}/J=(\Delta_c-\Delta)/\Delta_c$,
at $U/J$=1.26 and $n$=3.6. The measured rate (dots), is compared to the simplified model described in the text (squares).} \label{fig:gammadis}
\end{figure}

Let us now discuss in more detail the evolution of $\gamma$ in disordered lattices. We always
observe a rapid increase of $\gamma$ with increasing $\Delta$, as shown for example in
Fig.\ref{fig:gammadis}. As already discussed, the increase of $\gamma$ is fully justified by the
picture of a reduced effective tunnelling in presence of disorder. Theoretical models are however
available only for weakly-interacting systems in absence of a lattice \cite{Albert}, or for very
small momenta \cite{Khlebnikov}, therefore we cannot compare our data directly to theory. A simple
modification of the heuristic model used for the non-disordered case above to
$\gamma/2\pi=a\exp(-b\sqrt{nJ_{eff}/U})$, with a linear scaling of the effective tunnelling,
$J_{eff}=(\Delta_c-\Delta)/\Delta_c$, might capture the evolution of $\gamma(\Delta)$, for the same
parameters as before (see Fig.\ref{fig:gammadis}). The development of rigorous theoretical models
for the disordered case is clearly necessary.

\begin{figure}[htbp]
\includegraphics[width=0.9\columnwidth,clip]{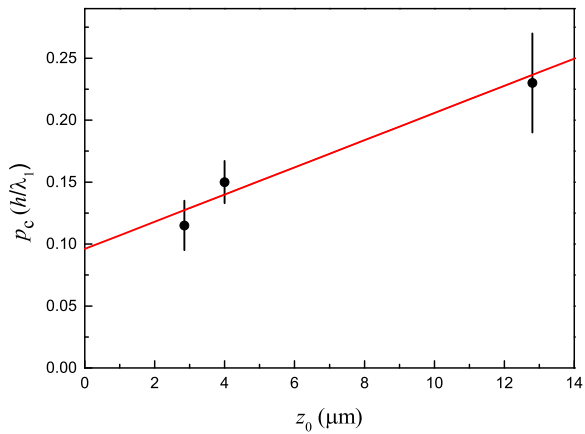}
\caption{Critical momentum vs the trap displacement, at $U/J$=1.26 and $\Delta/J$=5.
A linear extrapolation to zero displacement suggests a 30\% shift of $p_c$
with respect to the $z_0\rightarrow$0 limit.} \label{fig:shift}
\end{figure}

We have checked that the finite $z_0$ results only in a moderate shift of the critical momentum
$p_c$. In particular, for $U/J$=1.26 and $\Delta/J$=5, we measured $p_c$ for two other different
values of $z_0$ (Fig.\ref{fig:shift}). The observed downshift of $p_c$ with $z_0$ suggests that the
finite displacement, hence the finite force acting on the atoms, tends to shift the observed $p_c$
to higher values. This might be interpreted as a result of the finite response time of the
accelerating system to the increase of $\gamma(p)$. The effect is not big (about 30\% shift with
respect to the $z_0$=0 limit) and we expect that it should not change any of the conclusions of
this work.

\subsection{Fluid-insulator diagram}
The quasiperiodic lattice has a non-decaying correlation function \cite{Modugno}, which
implies a vanishing correlation energy $E_c$ on a system with infinite length. The experimental
system has however a relatively small length, and we can estimate an upper bound for $E_c$ from the
half-width of the first oscillation of the correlation function. This gives
$E_c\simeq8\hbar^2(\beta-1)^2/m\lambda_1^2\simeq J$, where $\beta=\lambda_1/\lambda_2$. For our
parameters, $E_c/J\simeq$0.7, which is almost one order of magnitude smaller of the explored range
of $E_{int}$. This puts the problem in the Thomas-Fermi limit of the mean-field theory
\cite{Savona09}.

\begin{figure}[htbp]
\includegraphics[width=0.9\columnwidth,clip]{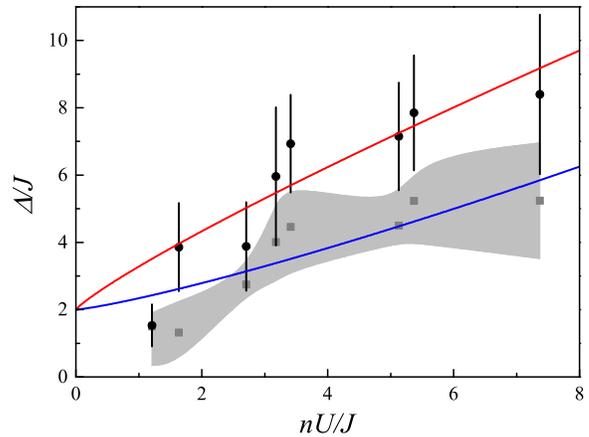}
\caption{(color online). Critical disorder to enter the insulating phase vs interaction energy.
The experimental data from the critical momentum (dots), excluding the first data point,
 are fitted with the superfluid-Bose glass model(red line). The experimental data for the center of the crossover
of $\delta p$ (squares), excluding the first two data points, are also fitted with the same model (blue line).
The grey area corresponds to an increase of $\delta p$ from
25\% to 75\% of its initial value.} \label{fig:diagram}
\end{figure}

It is interesting to compare the complementary information about the fluid-insulator crossover that
can be derived from the transport instability and from the broadening of the equilibrium momentum
distribution, i.e. from the evolution of the coherence of the system. In Fig.\ref{fig:diagram} we
compare the $\Delta_c$ data as in Fig.5 of the main paper to the characteristic values of $\Delta$
for the crossover in $\delta p$. The latter data were extracted from sigmoidal fits like the one
shown in Fig.4 of the main paper. The evolution of the two quantities is similar, but the coherence
data suggest an anticipated crossover from the fluid to the insulating regime. A fit of the
coherence data with the same scaling law $(\Delta_c-2)/J= A(NU/J)^\alpha$ gives $\alpha$=0.9(3) and
$A$=0.4(2), to be compared with $\alpha$=0.86(22) and $A$=1.3(4) already found for the transport
data.

These results on $\delta p$ can also be compared to a previous experiment we performed with the
same 1D quasiperiodic lattice, but with an essentially 3D Bose-Einstein condensate \cite{Deissler}.
By fitting the momentum-width data in \cite{Deissler} with the same scaling as above, we find again
an exponent close to unity ($\alpha$=1.0(1)), but a substantially larger coefficient $A=$6(1). The
comparison indicates that the presence of the radial degrees of freedom in the 3D system in
\cite{Deissler} was effectively reducing the disorder strength, therefore broadening the region of
existence of the superfluid phase in the $\Delta$-$E_{int}$ plane.

\end{document}